\documentclass[
    ,final            
    ,draft            
  ]
  {aipproc}

\layoutstyle{8x11single}

\date{}

\begin{document}

\title[\textit{Relic density and future colliders: inverse problem(s)}]{Relic density and future colliders: inverse problem(s)}


\author{Alexandre Arbey}{
  address={Universit\'e de Lyon, Lyon, F-69000, France; Universit\'e Lyon~1, Villeurbanne, \\
F-69622, France; Centre de Recherche Astrophysique de Lyon, Observatoire de Lyon, \\
9 avenue Charles Andr\'e, Saint-Genis Laval cedex, F-69561, France; CNRS, UMR 5574; \\
Ecole Normale Sup\'erieure de Lyon, Lyon, France.},
}

\author{Farvah Mahmoudi}{
  address={Laboratoire de Physique Corpusculaire de Clermont-Ferrand (LPC),\\ Universit\'e Blaise Pascal, CNRS/IN2P3, 63177 Aubi\`ere Cedex, France.}
}

\begin{abstract}
Relic density calculations are often used to constrain particle physics models, and in particular supersymmetry. We will show that the presence of additional energy or entropy before the Big-Bang nucleosynthesis can however completely change the relic density constraints on the SUSY parameter space. Therefore one should be extremely careful when using the relic density to constrain supersymmetry as it could give misleading results, especially if combined with the future collider data. Alternatively, we will also show that combining the discoveries of the future colliders with relic density calculations can shed light on the inaccessible pre-BBN dark time physics. Finally we will present SuperIso Relic, a new relic density calculator code in Supersymmetry, which incorporates alternative cosmological models, and is publicly available.
\end{abstract}

\maketitle


\section{Introduction}

The recent cosmological observations give evidence for the presence of a cosmological matter density, which represents about 27\% of the total density of the Universe, and the presence of the so-called dark energy. Using the total matter density observed by WMAP \cite{Komatsu:2008hk} and the baryon density indicated by Big-Bang nucleosynthesis (BBN) \cite{Burles:1997ez,Burles:1997fa} and including the theoretical uncertainties, the dark matter density range at 95\% C.L. is deduced \cite{Arbey:2008kv}:
\begin{equation}
 0.094 < \Omega_{DM} h^2 < 0.135 \;, \label{WMAPnew}
\end{equation}
where $h$ is the reduced Hubble constant. In the following, we also refer to the older range \cite{Ellis:1997wva}
\begin{equation}
 0.1 < \Omega_{DM} h^2 < 0.3 \;. \label{WMAPold}
\end{equation}
In supersymmetric models the lightest stable supersymmetric particle (LSP) constitutes the favorite candidate for dark matter. If the relic density can be calculated precisely, the accuracy of the latest WMAP data can therefore be used to constrain the supersymmetric parameters. The computation of the relic density is well-known within the standard model of cosmology \cite{Gondolo:1990dk,Edsjo:1997bg}, and is implemented in automatic codes, such as MicrOMEGAs \cite{Belanger:2006is}, DarkSUSY \cite{Gondolo:2004sc} or SuperIso Relic \cite{Arbey:2009gu}.

Nevertheless, the nature of the dark energy and the properties of the Universe in the pre-BBN epoch are still unknown, and the BBN era, at temperatures of about 1 MeV, is the oldest period in the cosmological evolution when reliable constraints are derived. The cosmology of the primordial Universe could therefore be much more complex, and the pre-BBN era could have for example experienced a slower or faster expansion. Such a modified expansion, even though still compatible with the BBN or the WMAP results, would change the LSP freeze-out time and the amount of relic density (see for example \cite{Arbey:2008kv,Kamionkowski:1990ni,Salati:2002md,Chung:2007cn}). A similar question exists concerning the entropy content of the Universe at that period, and the eventual consequences on the energy conservation in the primodial epochs (see for example \cite{Moroi:1999zb,Giudice:2000ex,Gelmini:2006pw,Arbey:2009gt}).

In the following, we will first describe the way the relic density is calculated in the standard cosmology. We will then present how the calculations can be modified in altered cosmological scenarios. We then analyze the consequences of the cosmological uncertainties on the supersymmetric parameter searches. We also inverse the problem and show that the determination of a beyond the Standard Model particle physics scenario will give hints on the cosmological properties of the early Universe. Finally, we will present the SuperIso Relic package and conclude.

\section{Relic density in Standard Cosmology}
The cosmological standard model is based on a Friedmann-Lema{\^\i}tre Universe, approximately flat, incorporating a cosmological constant accelerating its expansion, and filled with radiation, baryonic matter and cold dark matter. Before BBN, the Universe expansion is dominated by the radiation density, and therefore the expansion rate $H$ of the Universe is determined by the Friedmann equation
\begin{equation}
H^2=\frac{8 \pi G}{3} \rho_{rad}\;,\label{friedmann_stand}
\end{equation}
where
\begin{equation}
\rho_{rad}(T)=g_{\mbox{eff}}(T) \frac{\pi^2}{30} T^4
\end{equation}
is the radiation density and $g_{\mbox{eff}}$ is the effective number of degrees of freedom of radiation. The computation of the relic density is based on the solution of the Boltzmann evolution equation \cite{Gondolo:1990dk,Edsjo:1997bg}
\begin{equation}
dn/dt=-3Hn-\langle \sigma_{\mbox{eff}} v\rangle (n^2 - n_{\mbox{eq}}^2)\;, \label{evol_eq}
\end{equation}
where $n$ is the number density of all supersymmetric particles, $n_{\mbox{eq}}$ their equilibrium density, and $\langle \sigma_{\mbox{eff}} v\rangle$ is the thermal average of the annihilation rate of the supersymmetric particles to the Standard Model particles. By solving this equation, the density number of supersymmetric particles in the present Universe and consequently the relic density can be determined.

The computation of the thermally averaged annihilation cross section $\langle \sigma_{\mbox{eff}} v \rangle$ requires the computation of the many annihilation and co-annihilation amplitudes. The annihilation rate of supersymmetric particles $i$ and $j$ into SM particles $k$ and $l$ is defined as \cite{Gondolo:1990dk,Edsjo:1997bg}:
\begin{equation}
W_{ij\to kl} = \frac{p_{kl}}{16\pi^2 g_i g_j S_{kl} \sqrt{s}} \sum_{\rm{internal~d.o.f.}} \int \left| \mathcal{M}(ij\to kl) \right|^2 d\Omega \;, \label{Weff}
\end{equation}
where $\mathcal{M}$ is the transition amplitude, $s$ the center-of-mass energy, $g_i$ the number of degrees of freedom of the particle $i$ and $p_{kl}$ the final center-of-mass momentum such as
\begin{equation}
p_{kl} = \frac{\left[s-(m_k+m_l)^2\right]^{1/2} \left[s-(m_k-m_l)^2\right]^{1/2}}{2\sqrt{s}}\;.
\end{equation}
$S_{kl}$ in Eq. (\ref{Weff}) is a symmetry factor equal to 2 for identical final particles and to 1 otherwise, and the integration is over the outgoing directions of one of the final particles. Moreover, an average over initial internal degrees of freedom is performed.

The effective annihilation rate $W_{\rm eff}$ can be defined by
\begin{equation}
g_{LSP}^2 p_{\rm{eff}} W_{\rm{eff}} \equiv \sum_{ij} g_i g_j p_{ij} W_{ij}
\end{equation}
with
\begin{equation}
p_{\rm{eff}}(\sqrt{s}) = \frac{1}{2} \sqrt{(\sqrt{s})^2 -4 m_{LSP}^2} \;.
\end{equation}
The following relation can therefore be deduced:
\begin{equation}
\frac{d W_{\rm eff}}{d \cos\theta} = \sum_{ijkl} \frac{p_{ij} p_{kl}}{ 8 \pi g_{LSP}^2 p_{\rm eff} S_{kl} \sqrt{s} }
\sum_{\rm helicities} \left| \sum_{\rm diagrams}  \mathcal{M}(ij \to kl) \right|^2 \;,
\end{equation}
where $\theta$ is the angle between particles $i$ and $k$. The thermal average of the effective cross section is then obtained by:
\begin{equation}
\langle \sigma_{\rm{eff}}v \rangle = \displaystyle\frac{\displaystyle\int_0^\infty dp_{\rm{eff}} p_{\rm{eff}}^2 W_{\rm{eff}}(\sqrt{s}) K_1 \left(\displaystyle\frac{\sqrt{s}}{T} \right) } { m_{LSP}^4 T \left[ \displaystyle\sum_i \displaystyle\frac{g_i}{g_{LSP}} \displaystyle\frac{m_i^2}{m_1^2} K_2 \left(\displaystyle\frac{m_i}{T}\right) \right]^2}\;,
\end{equation}
where $K_1$ and $K_2$ are the modified Bessel functions of the second kind of order 1 and 2 respectively.

The ratio of the number density to the radiation entropy density $Y(T)=n(T)/s(T)$ is defined, in which
\begin{equation}
s(T)=h_{\mbox{eff}}(T) \frac{2 \pi^2}{45} T^3 \;.
\end{equation}
$h_{\mbox{eff}}$ is the effective number of entropic degrees of freedom of radiation. Combining Eqs. (\ref{friedmann_stand}) and (\ref{evol_eq}) and defining the ratio of the LSP mass over temperature $x=m_{\mbox{\small LSP}}/T$, yield
\begin{equation}
\frac{dY}{dx}=-\sqrt{\frac{\pi}{45 G}}\frac{g_*^{1/2} m_{\mbox{\small LSP}}}{x^2} \langle \sigma_{\mbox{eff}} v\rangle (Y^2 - Y^2_{\mbox{eq}}) \;, \label{main}
\end{equation}
with
\begin{equation}
g_*^{1/2}=\frac{h_{\mbox{eff}}}{\sqrt{g_{\mbox{eff}}}}\left(1+\frac{T}{3 h_{\mbox{eff}}}\frac{dh_{\mbox{eff}}}{dT}\right) \;,
\end{equation}
and
\begin{equation}
Y_{eq} = \frac{45}{4 \pi^4 T^2} h_{\rm{eff}} \sum_i g_i m_i^2 K_2\left(\frac{m_i}{T}\right) \;,
\end{equation}
where $i$ runs over all supersymmetric particles of mass $m_i$ and with $g_i$ degrees of freedom. The freeze-out temperature $T_f$ is the temperature at which the LSP leaves the initial thermal equilibrium when $Y (T_f) = (1 + \delta) Y_{\mbox{eq}}(T_f)$, with $\delta \simeq 1.5$. The relic density is obtained by integrating Eq. (\ref{main}) from $x=0$ to $m_{\mbox{\small LSP}}/T_0$, where $T_0=2.726$ K is the temperature of the Universe today \cite{Gondolo:1990dk,Edsjo:1997bg}:
\begin{equation}
\Omega_{\mbox{\small LSP}} h^2 = \frac{m_{\mbox{\small LSP}} s(T_0) Y(T_0) h^2}{\rho_c^0} \approx 2.755\times 10^8 \frac{m_{\mbox{\small LSP}}}{1 \mbox{ GeV}} Y(T_0)\;,
\end{equation}
where $\rho_c^0$ is the critical density of the Universe, such as
\begin{equation}
H^2_0 = \frac{8 \pi G}{3} \rho_c^0 \;,
\end{equation}
$H_0$ being the Hubble constant.
\section{Relic density in Alternative Cosmological Scenarios}
In presence of non-thermal production of SUSY particles, the Boltzmann equation becomes
\begin{equation}
\frac{dn}{dt} = - 3 H n - \langle \sigma v \rangle (n^2 - n^2_{eq}) + N_D  \;.\label{boltzmann}
\end{equation}
The term $N_D$ is added to provide a parametrization of the non-thermal production of SUSY particles. The expansion rate $H$ can also be modified: following \cite{Arbey:2008kv,Arbey:2009gt}, $\rho_D$ is introduced as an effective dark density which parametrizes the expansion rate modification. The Friedmann equation then becomes
\begin{equation}
 H^2=\frac{8 \pi G}{3} (\rho_{rad} + \rho_D)  \;,\label{friedmann}
\end{equation}
where $\rho_{rad}$ is the radiation energy density, which is considered as dominant before BBN in the standard cosmological model. In case of additional entropy fluctuations, the entropy evolution reads
\begin{equation}
\frac{ds}{dt} = - 3 H s + \Sigma_D \label{entropy_evolution} \;,
\end{equation}
where $s$ is the total entropy density. $\Sigma_D$ parametrizes here effective entropy fluctuations due to unknown properties of the early Universe.

Separating the radiation entropy density from the total entropy density, {\it i.e.} setting $s \equiv s_{rad} + s_D$ where $s_{rad}$ is the radiation entropy density and $s_D$ is an effective entropy density, the following relation between $s_D$ and $\Sigma_D$ can be derived:
\begin{equation}
\Sigma_D = \sqrt{\frac{4 \pi^3 G}{5}} \sqrt{1 + \tilde{\rho}_D} T^2 \left[\sqrt{g_{\rm{eff}}} s_D - \frac13  \frac{h_{\rm{eff}}}{g_*^{1/2}} T \frac{ds_D}{dT}\right] \;.
\end{equation}
Following the standard relic density calculation method \cite{Gondolo:1990dk,Edsjo:1997bg}, we introduce $Y \equiv n/s$, and Eq. (\ref{boltzmann}) becomes
\begin{equation}
 \frac{dY}{dx}= - \frac{m_{LSP}}{x^2} \sqrt{\frac{\pi}{45 G}} g_*^{1/2} \left( \frac{1 + \tilde{s}_D}{\sqrt{1+\tilde{\rho}_D}} \right) \left[\langle \sigma v \rangle (Y^2 - Y^2_{eq}) + \frac{Y \Sigma_D - N_D}{\left(h_{\rm{eff}}(T) \frac{2\pi^2}{45} T^3\right)^2 (1+\tilde{s}_D)^2} \right] \;, \label{final}
\end{equation}
where $x=m_{LSP}/T$, $m_{LSP}$ being the mass of the relic particle,
\begin{equation}
 \tilde{s}_D \equiv \frac{s_D}{h_{\rm{eff}}(T) \frac{2\pi^2}{45} T^3}\;, \qquad\qquad \tilde{\rho}_D \equiv \frac{\rho_D}{g_{\rm{eff}} \frac{\pi^2}{30} T^4}\;,
\end{equation}
and
\begin{equation}
 Y_{eq} = \frac{45}{4 \pi^4 T^2} h_{\rm{eff}} \frac{1}{(1+\tilde{s}_D)} \sum_i g_i m_i^2 K_2\left(\frac{m_i}{T}\right) \;.
\end{equation}
The relic density can then be calculated in the standard way:
\begin{equation}
 \Omega h^2 = 2.755 \times 10^8 Y_0 m_{LSP}/\mbox{GeV} \;.
\end{equation}
where $Y_0$ is the present value of $Y$. In the limit where $\rho_D = s_D = \Sigma_D = N_D = 0$, usual relations are retrieved. We should note here that $s_D$ and $\Sigma_D$ are not independent variables. In the following, we neglect $N_D$.

We use the parametrizations described in \cite{Arbey:2008kv,Arbey:2009gt} for $\rho_D$ and $s_D$:
\begin{equation}
 \rho_D =  \kappa_\rho \rho_{rad}(T_{BBN}) \left(\frac{T}{T_{BBN}}\right)^{n_\rho} \;,\label{rhoD}
\end{equation}and
\begin{equation}
 s_D =  \kappa_s s_{rad}(T_{BBN}) \left(\frac{T}{T_{BBN}}\right)^{n_s} \;,\label{sD}
\end{equation}
where $T_{BBN}$ is the BBN temperature. $\kappa_\rho$ ($\kappa_s$) is the ratio of effective dark energy (entropy) density over radiation energy (entropy) density at BBN time, and $n_\rho$ and $n_s$ are parameters describing the behavior of the densities. We refer to \cite{Arbey:2008kv,Arbey:2009gt} for detailed descriptions and discussions on these parametrizations.
\section{Supersymmetric consequences}
\begin{figure}[t!]
$\begin{array}{cc}
\includegraphics[width=6.05cm]{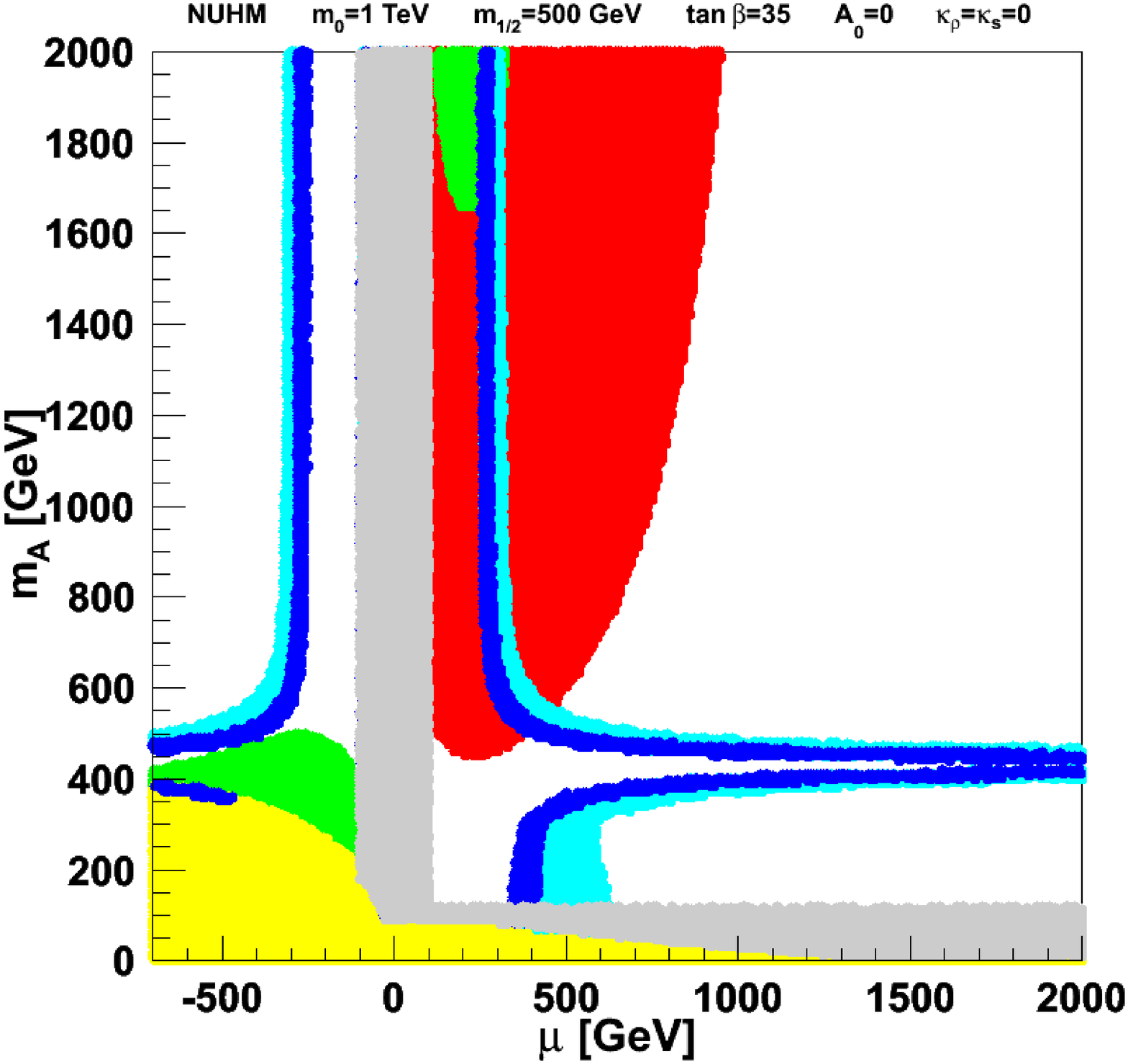}&\includegraphics[width=6.05cm]{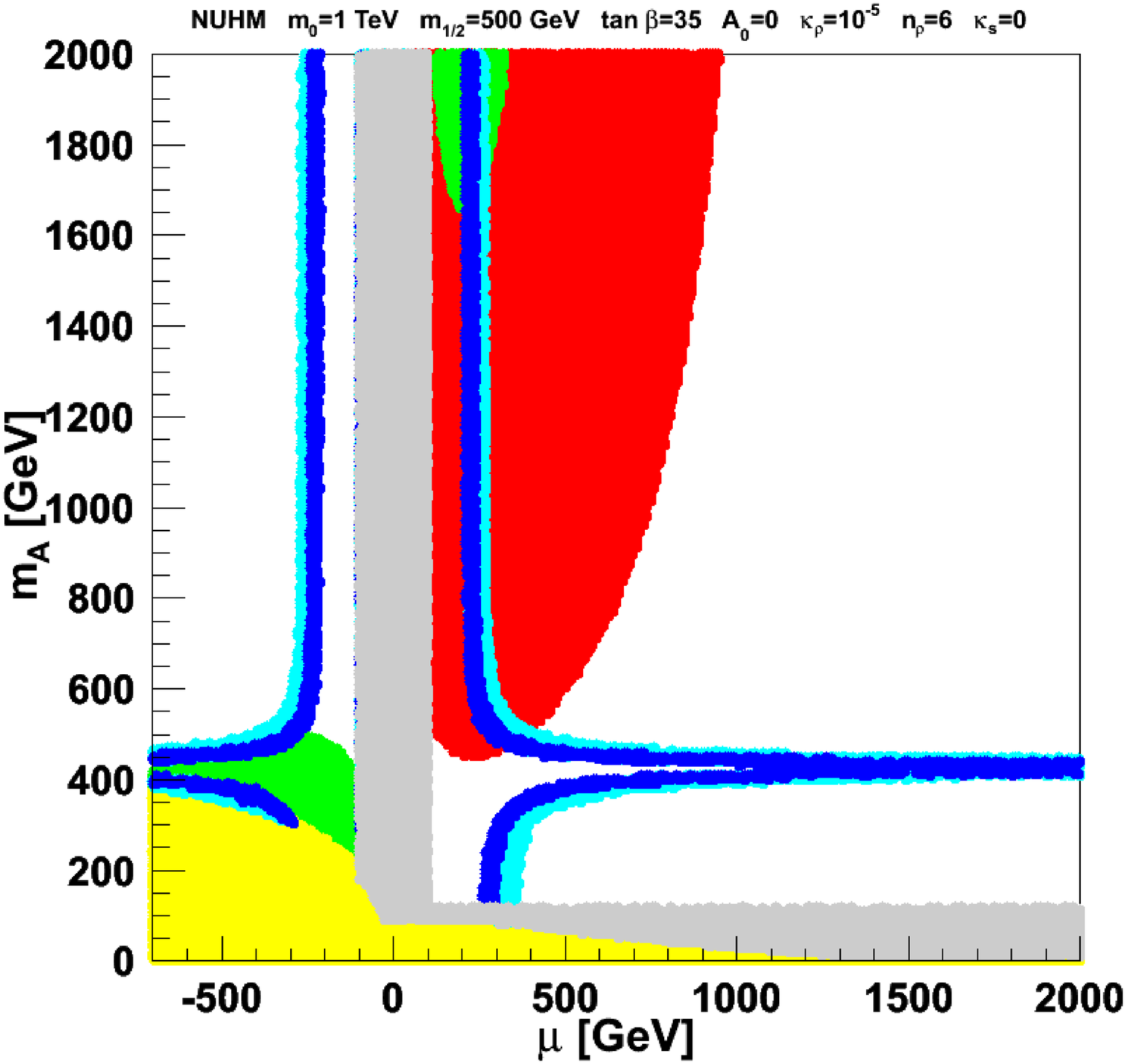}\\
\includegraphics[width=6.05cm]{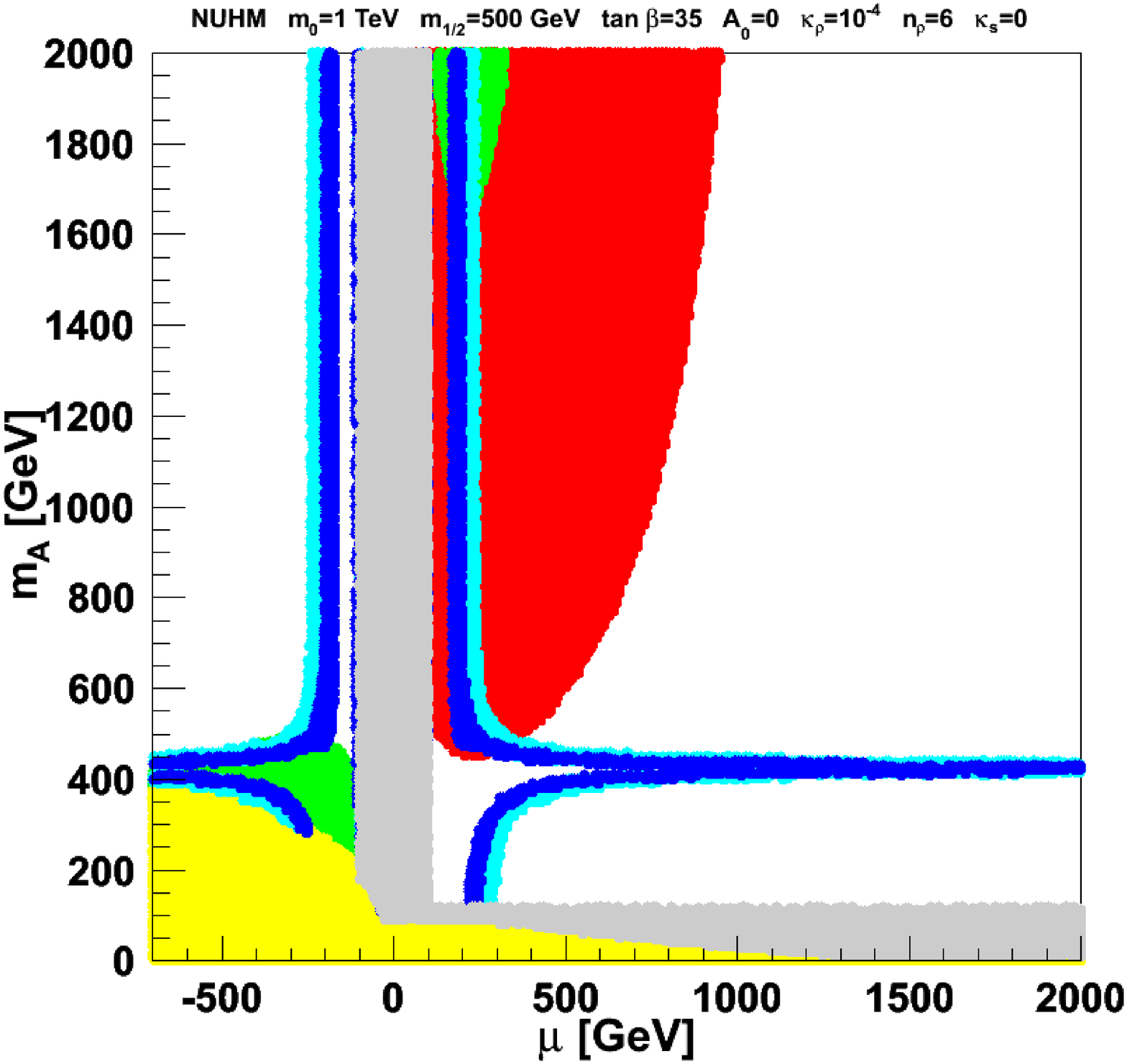}&\includegraphics[width=6.05cm]{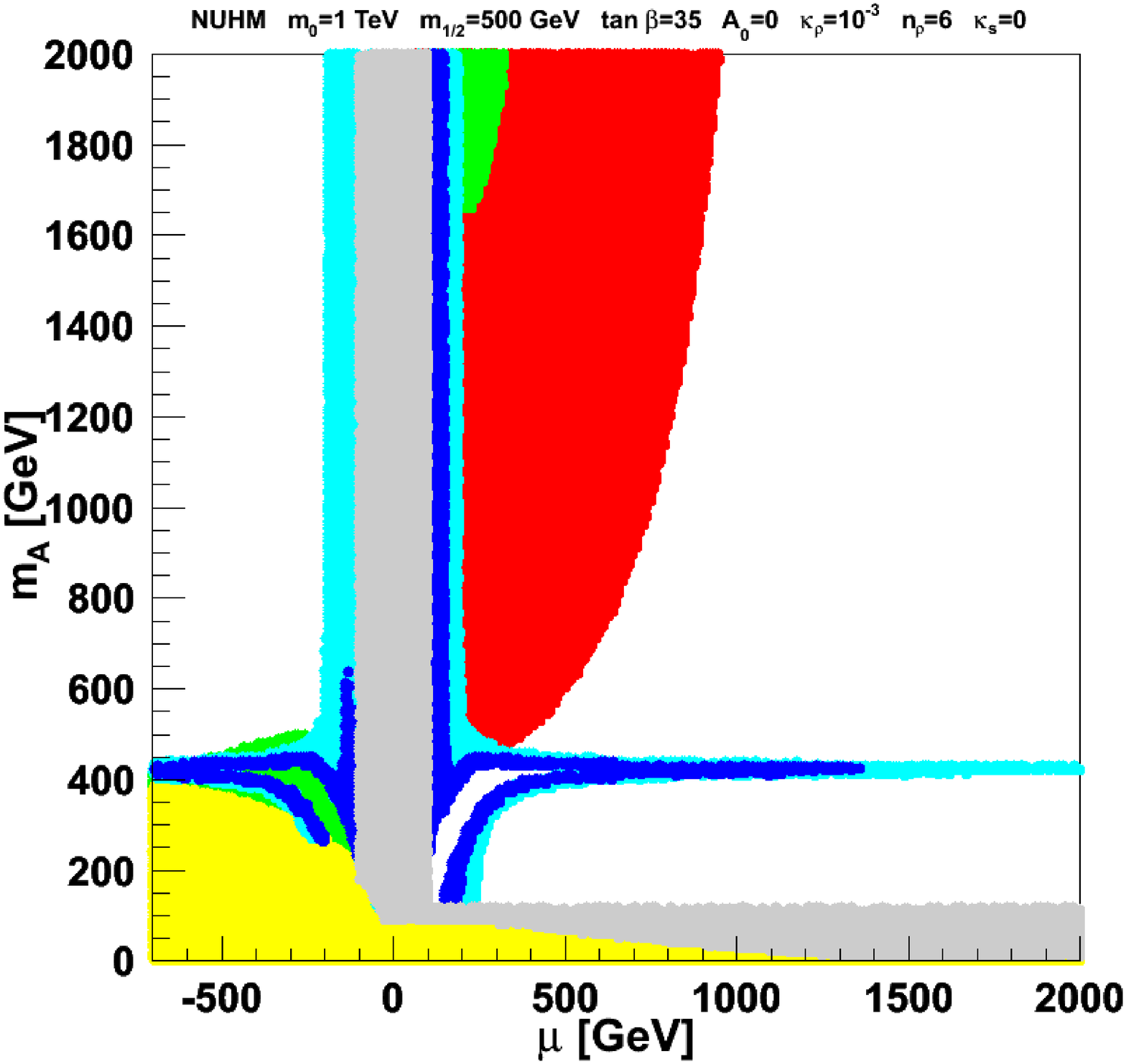}\\
\includegraphics[width=6.05cm]{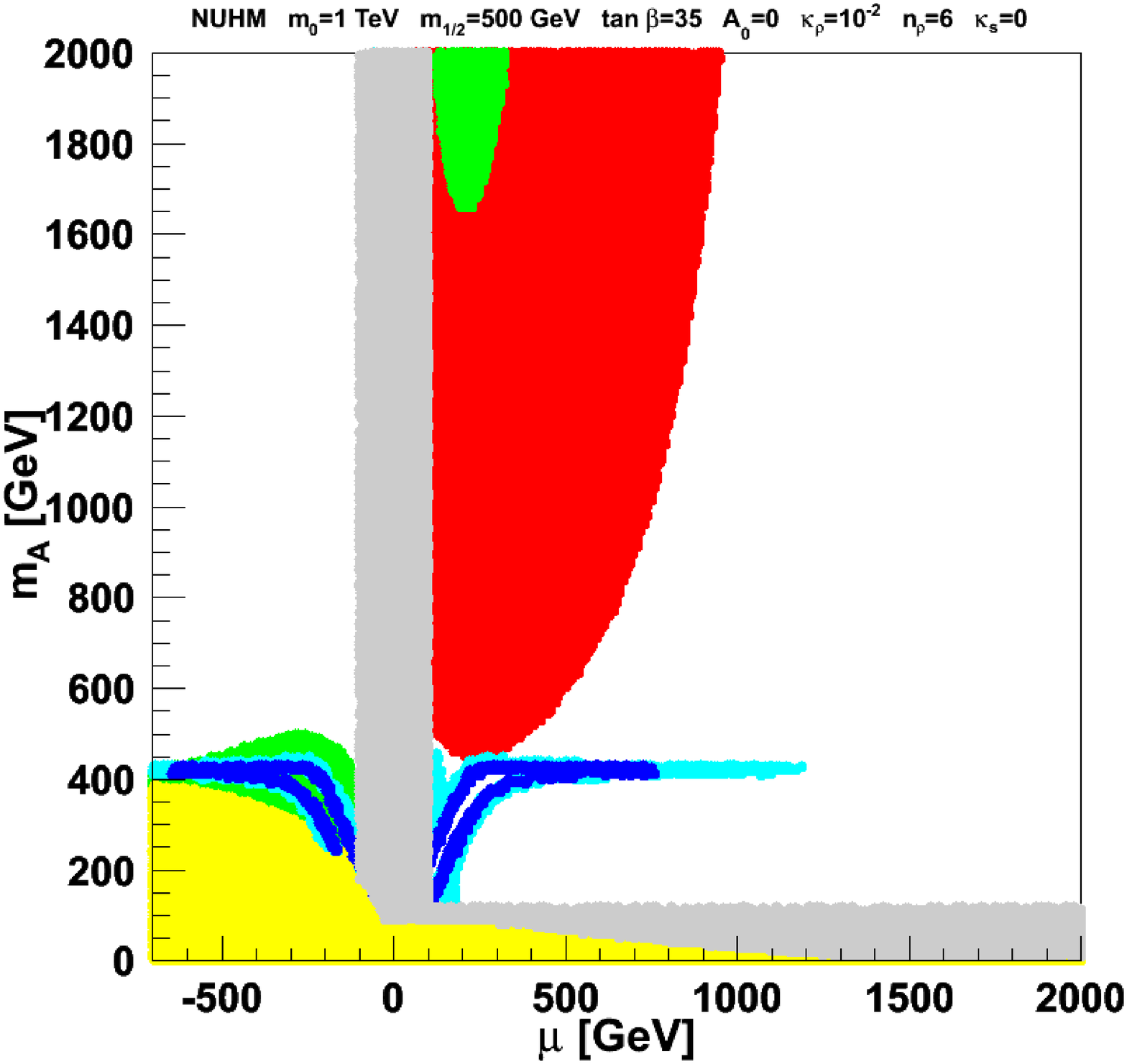}&\includegraphics[width=6.05cm]{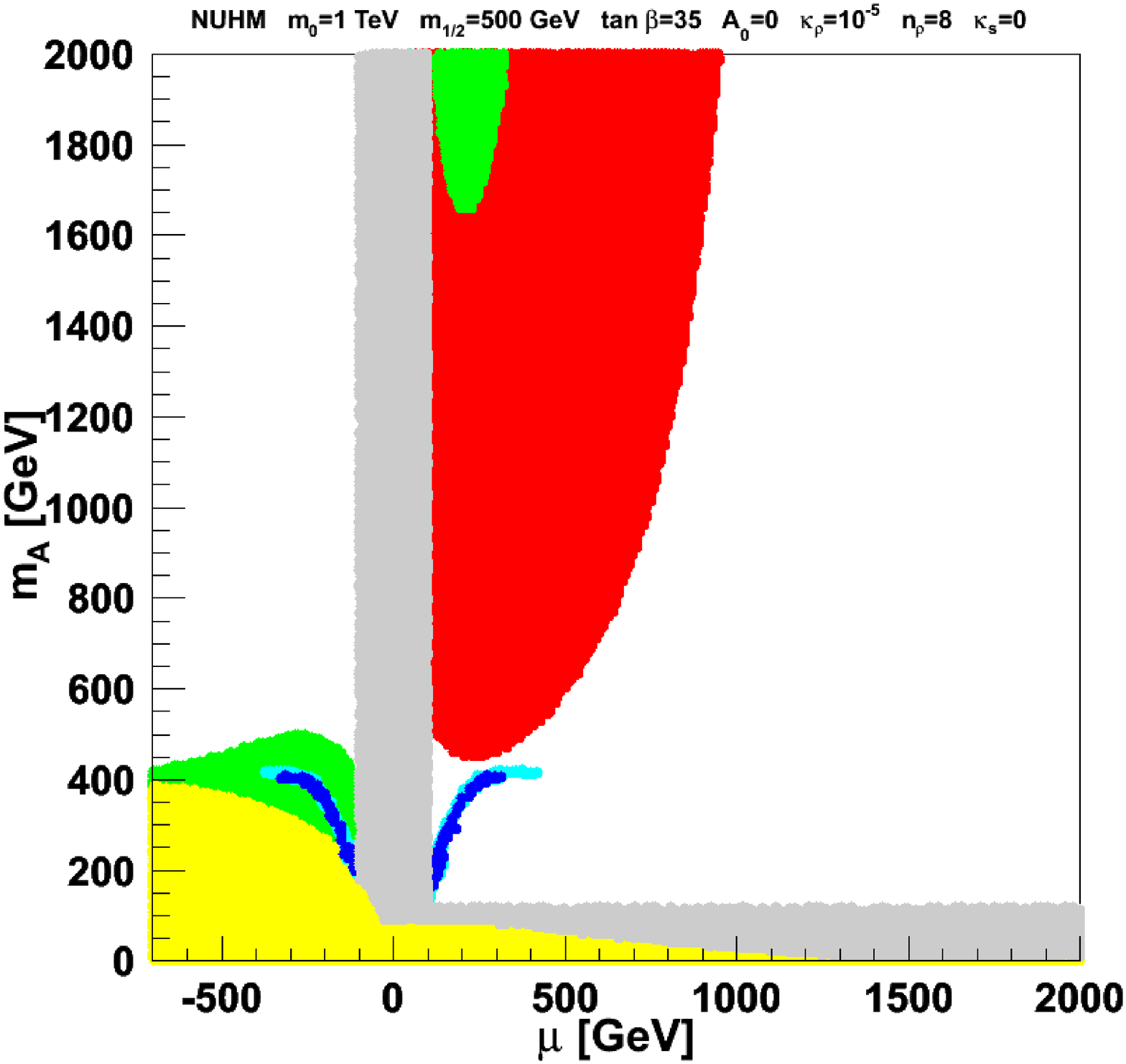}
\end{array}$
\caption{Constraints in the NUHM parameter plane $(\mu,m_A)$ in presence of additional dark energy for several values of $\kappa_\rho$ and $n_\rho$, and for $m_0=1$ TeV, $m_{1/2}=500$ GeV, $\tan\beta=35$, $A_0=0$. The color code is given in the text.\label{NUHMenergy}}
\end{figure}

Relic density is often used to constrain SUSY parameter space (see for example \cite{Battaglia:2003ab}). In particular the non-universal Higgs model (NUHM) provides attractive candidates for dark matter \cite{Ellis:2007ka}. For our analysis, we consider the NUHM parameter plane $(\mu,m_A)$, fixing the other parameters ($m_0=1$ TeV, $m_{1/2}=500$ GeV, $\tan\beta=35$, $A_0=0$). About 250,000 random SUSY points in the NUHM parameter plane ($\mu$,$m_A$) are generated using SOFTSUSY v2.0.18 \cite{softsusy}, and for each point we compute flavor physics observables, direct limits and the relic density with SuperIso Relic v2.7 \cite{Arbey:2009gu}. In Fig.~\ref{NUHMenergy}, the excluded zones due to different observables are displayed: the red area is excluded by the isospin asymmetry of $B \to K^* \gamma$, the green by the inclusive branching ratio of $b \to s \gamma$, the yellow area leads to tachyonic particles and the gray zone has been excluded by collider searches. All these exclusions are related to particle physics and are subject to uncertainties which are under control. The dark (light) blue zones are \emph{favored} by the WMAP (old) dark matter constraints. Hence, in the top left figure for example, which corresponds to the standard cosmological model, only tiny strips remain compatible with all the displayed constraints, and the relic density observable happens to be extremely constraining. This kind of figures is often used to determine the favored SUSY parameter space zones.\\
In the next plots of Fig.~\ref{NUHMenergy} however, we show that the presence of additional energy density in the early Universe, even negligible at BBN time, can completely change the results. These plots show the influence of a quintessence--like dark energy ($n_\rho=6$) whose proportions relative to the radiation density at BBN time are respectively $\kappa_\rho=10^{-5}$, $10^{-4}$, $10^{-3}$ and $10^{-2}$. The last plot shows the influence of a density of a decaying scalar field ($n_\rho=8$) with an extremely low $\kappa_\rho=10^{-5}$. In these plots the relic favored zones are moved towards the center, completely modifying the favored SUSY parameters.

\begin{figure}[t!]
$\begin{array}{cc}
\includegraphics[width=6.05cm]{plot1.eps}&\includegraphics[width=6.05cm]{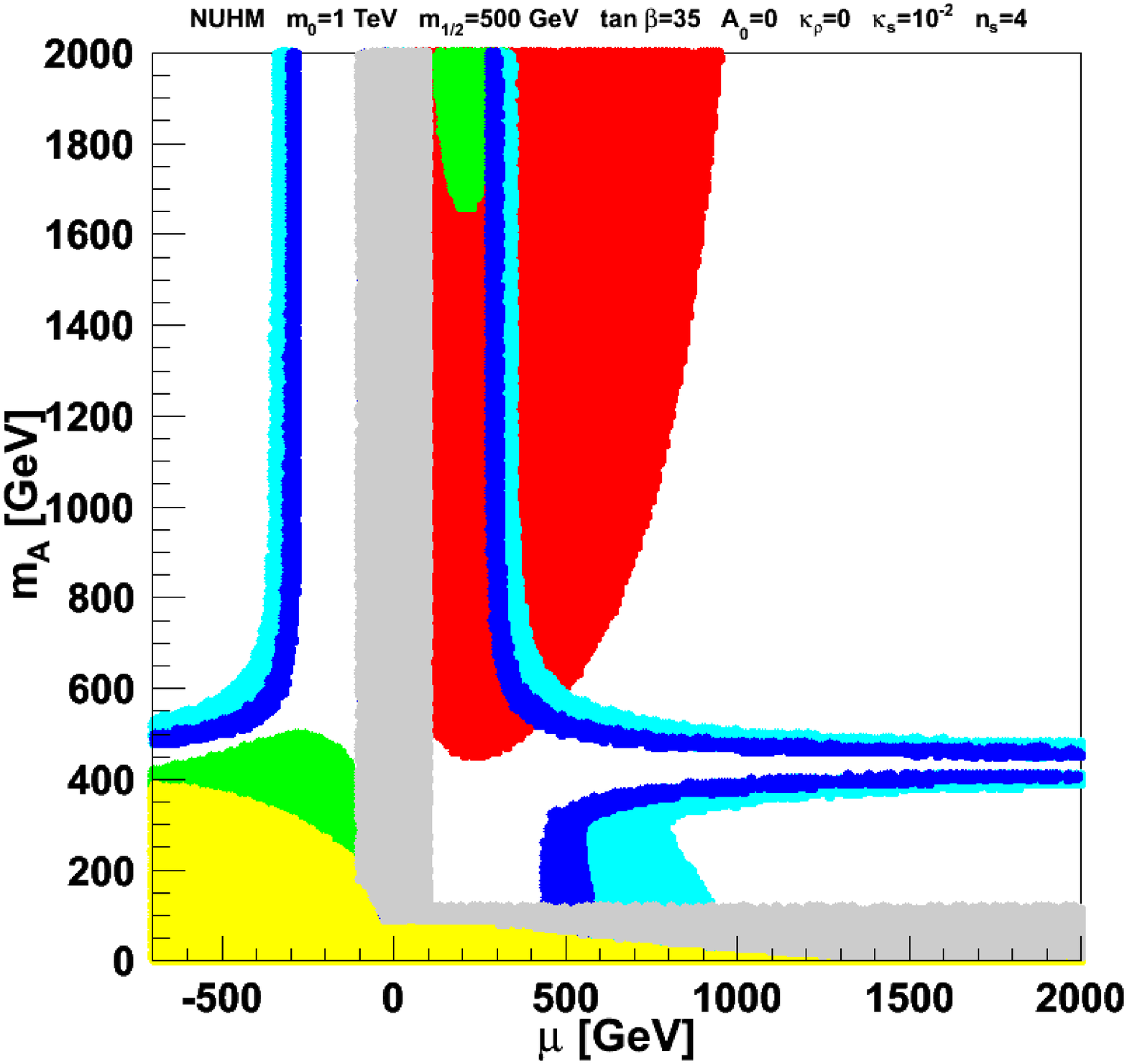}\\
\includegraphics[width=6.05cm]{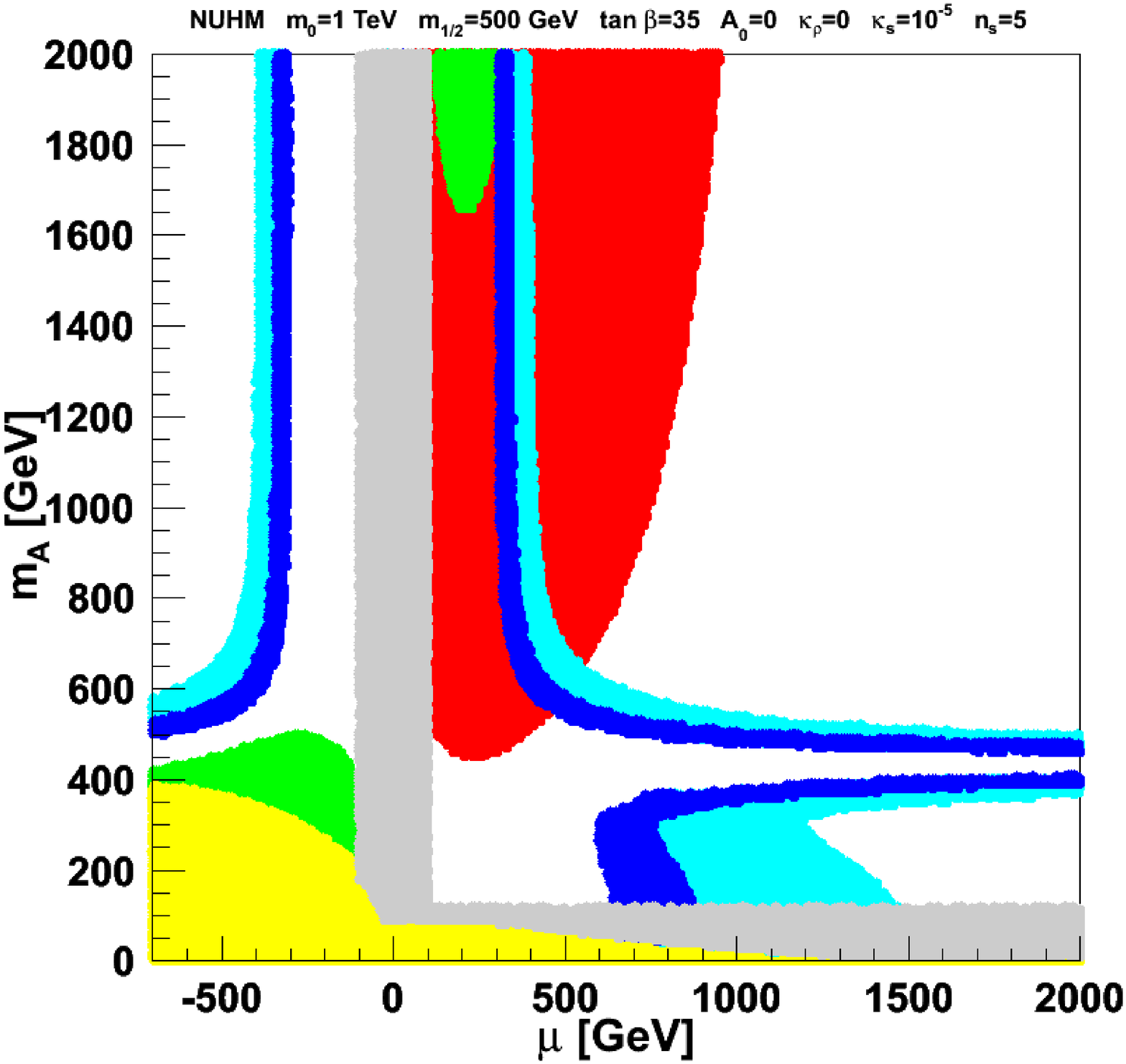}&\includegraphics[width=6.05cm]{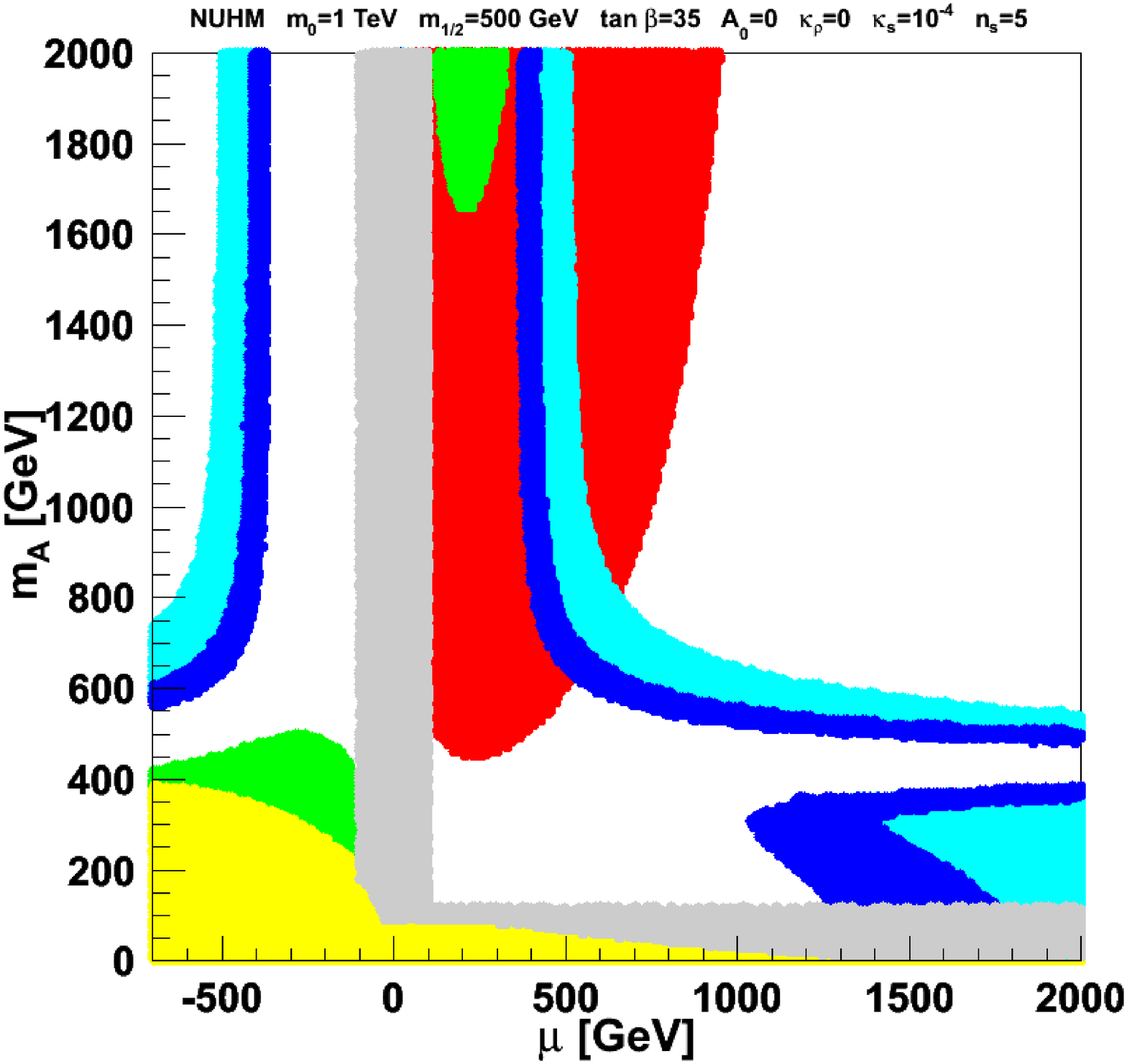}\\
\includegraphics[width=6.05cm]{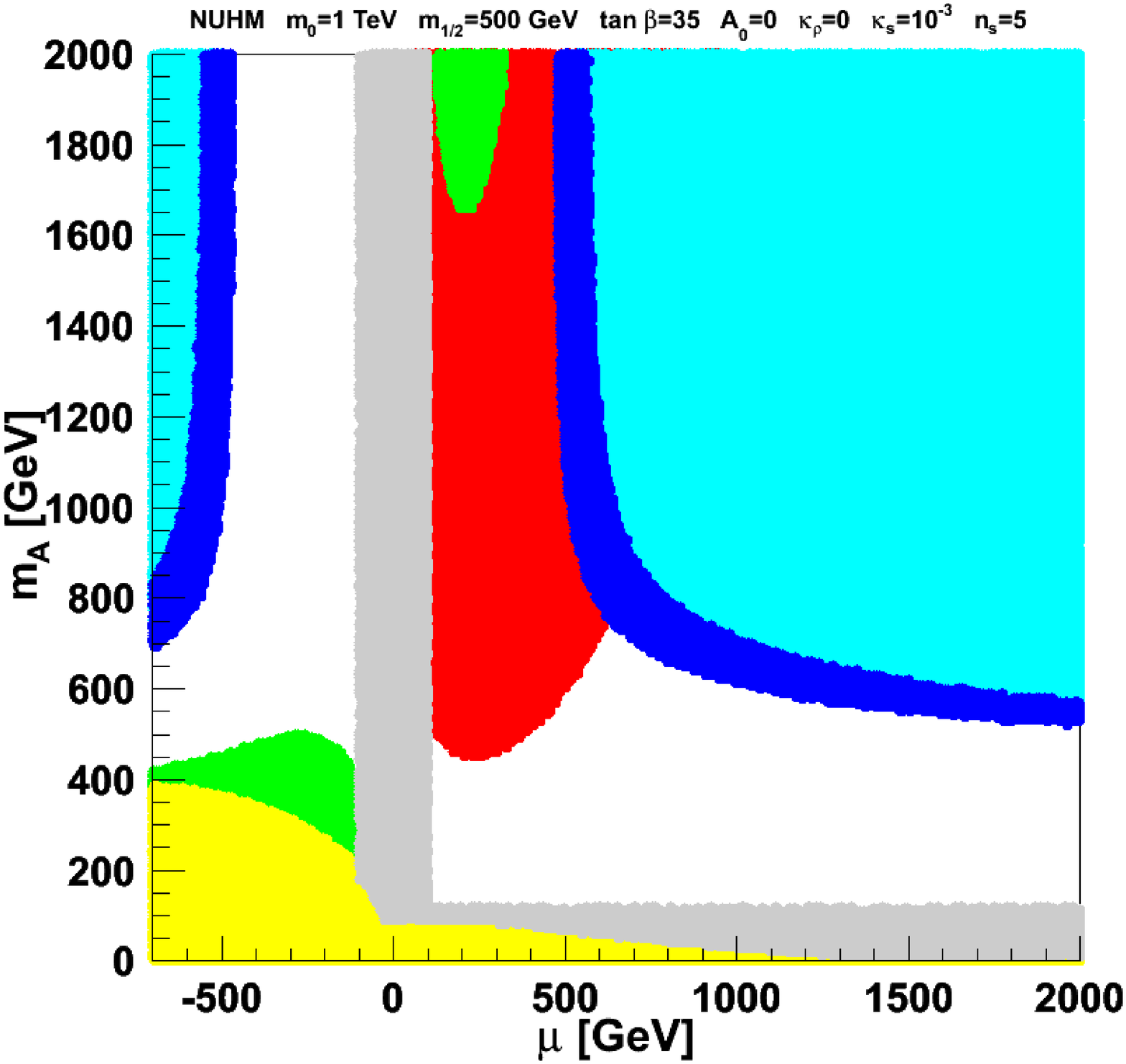}&\includegraphics[width=6.05cm]{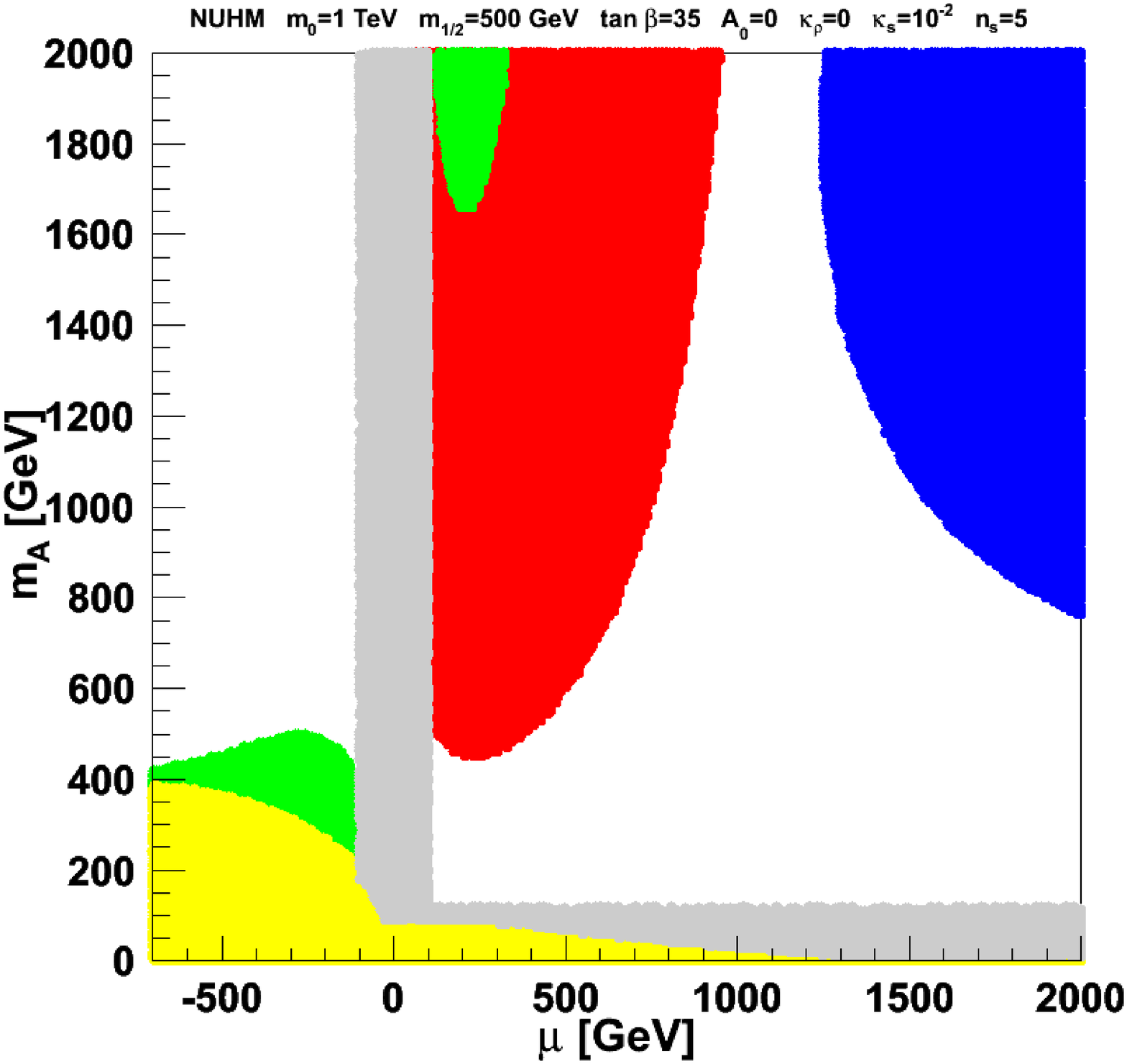}\\
\end{array}$
\caption{Constraints in the NUHM parameter plane $(\mu,m_A)$ in presence of additional dark energy for several values of $\kappa_s$ and $n_s$, and for $m_0=1$ TeV, $m_{1/2}=500$ GeV, $\tan\beta=35$, $A_0=0$. The color code is given in the text.\label{NUHMentropy}}
\end{figure}%
The result is similar when considering the influence of additional entropy. In Fig.~\ref{NUHMentropy}, the first plot shows for reference the constraints in the standard model of cosmology. The second plot presents the influence of a dark entropy density with $n_s=4$ and $\kappa_s=10^{-3}$, which can occur in the case of reheating. The next plots show the influence of a dark entropy density with $n_s=5$ and $\kappa_s=10^{-4}$, $\kappa_s=10^{-3}$ and $\kappa_s=10^{-2}$. The relic density favored areas are this time moved outwards with different shapes.

From these two analyses, it is clear that the relic density calculations can be strongly altered by the presence of entropy or energy densities, even very small or negligible at the BBN time ({\it i.e.} completely unobservable in the current cosmological data). Thus, cosmological unknown properties of the early Universe can completely change the favored SUSY parameter space, and could lead to erroneous affirmations on the properties of the SUSY particles.
\section{Inverse problem}
We have seen in the previous section that it is critical to use the relic density to constrain SUSY. However the problem can be inverted: the future colliders will give the possibility to determine the new physics  and relic particle properties \cite{Baltz:2006fm}, and by calculating the relic density in different cosmological scenarios it would be possible to determine or to verify some of the physical properties of the early Universe. In such cases, combining the relic density constraints with the BBN limits would remove some of the degeneracies.
\begin{figure}[t!]
\includegraphics[width=6.05cm]{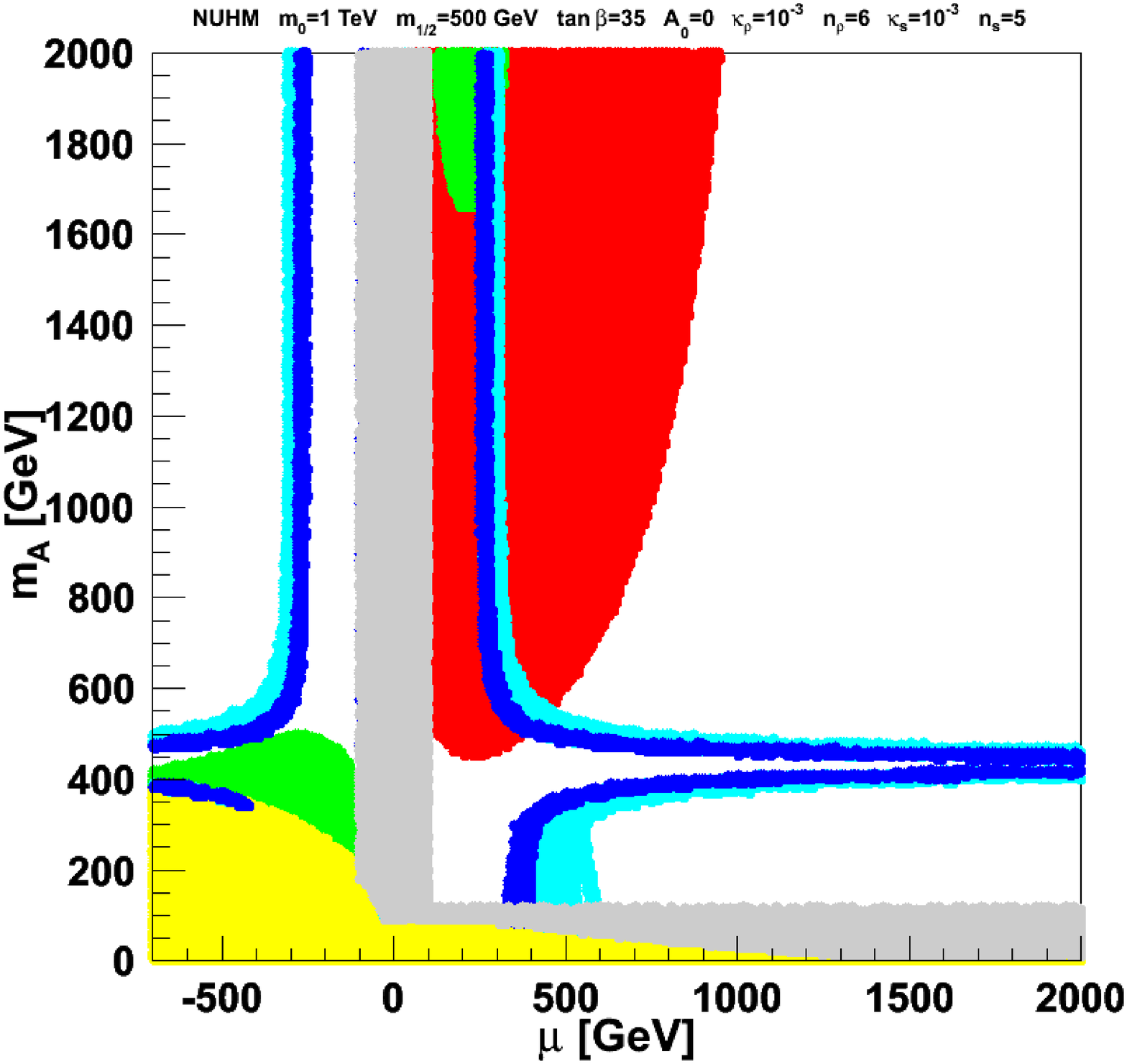}\includegraphics[width=6.05cm]{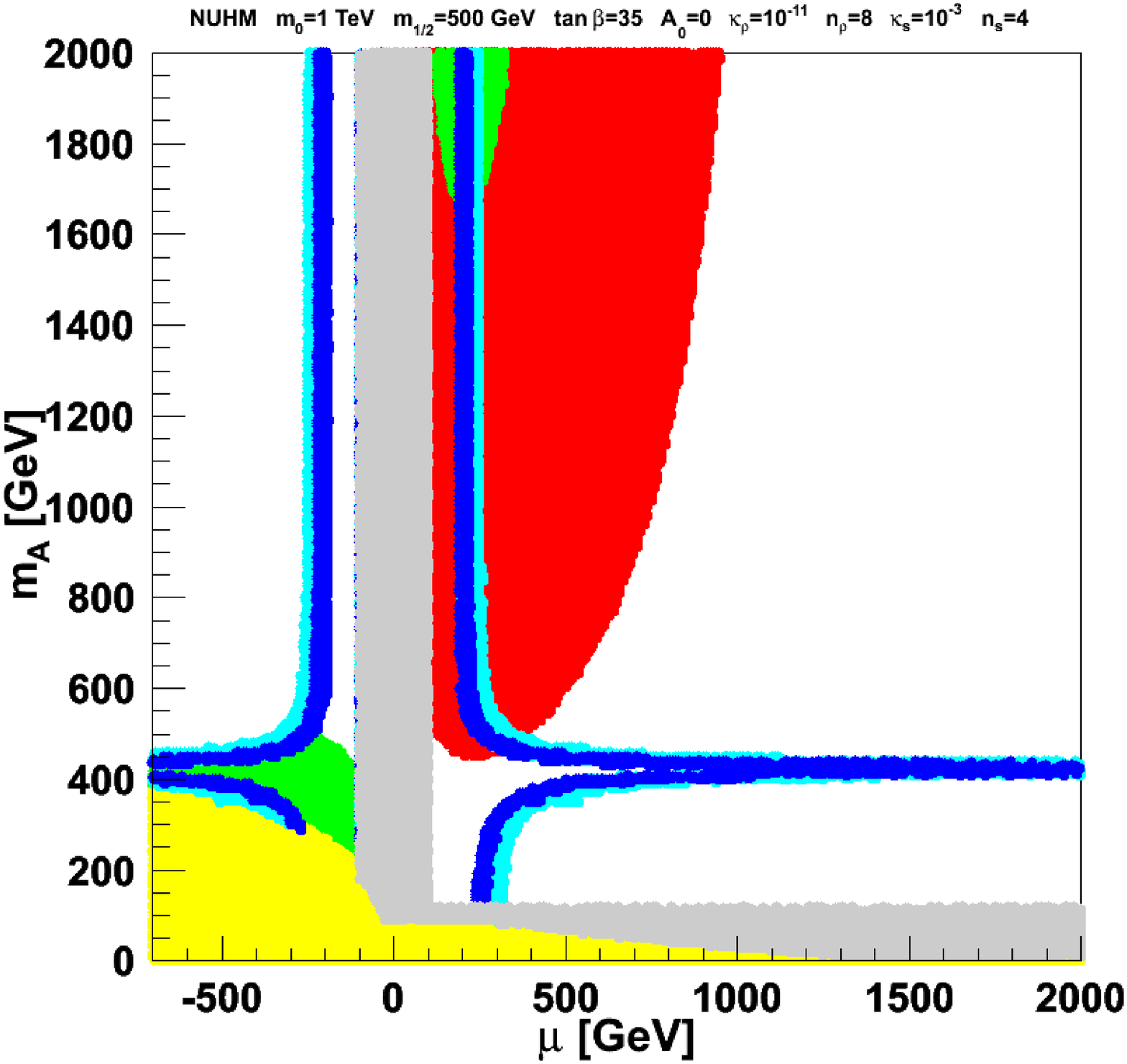}\\
\caption{Constraints in the NUHM parameter plane $(\mu,m_A)$ in presence of additional dark entropy and energy for $\kappa_\rho=10^{-2}$, $n_\rho=6$, $\kappa_s=10^{-2}$, $n_s=5$ (left) and $\kappa_\rho=10^{-11}$, $n_\rho=8$, $\kappa_s=10^{-2}$, $n_s=4$ (right). The color code is given in the text.\label{degeneracy}}
\end{figure}%
In Fig.~\ref{degeneracy} for example, the constraints are presented for two different cosmological scenarios: $\kappa_\rho=10^{-2}$, $n_\rho=6$, $\kappa_s=10^{-2}$, $n_s=5$ on the left, $\kappa_\rho=10^{-11}$, $n_\rho=8$, $\kappa_s=10^{-2}$, $n_s=4$ on the right. The cosmological properties are different, but the obtained zones are rather similar, which means that there is a degeneracy between dark energy and dark entropy effects. Let us assume that the subjacent SUSY model leads to a relic density inside the WMAP favored zone of these plots, and disfavored in the standard cosmology model. 
\begin{figure}[t!]
\includegraphics[width=6.05cm]{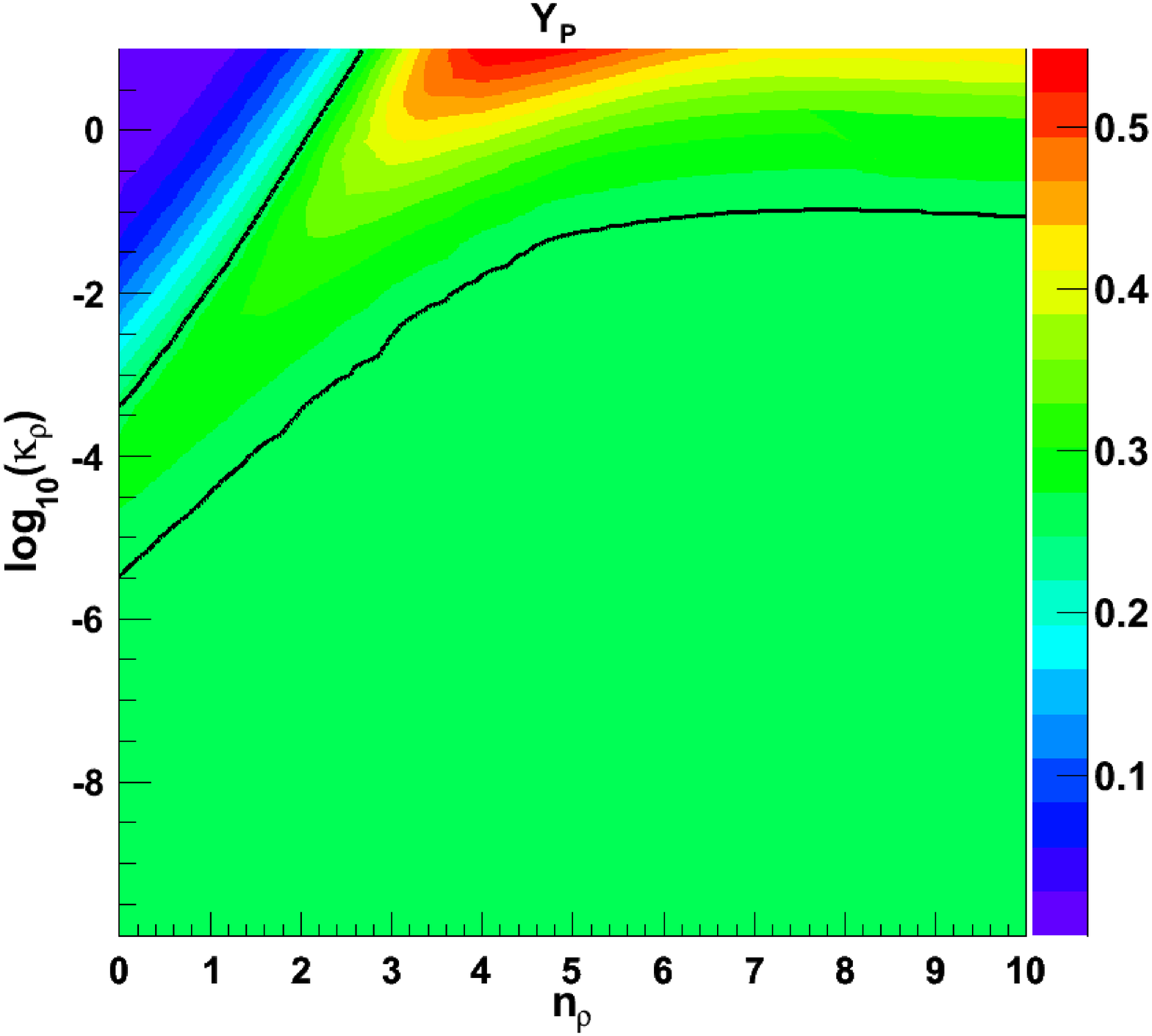}\includegraphics[width=6.05cm]{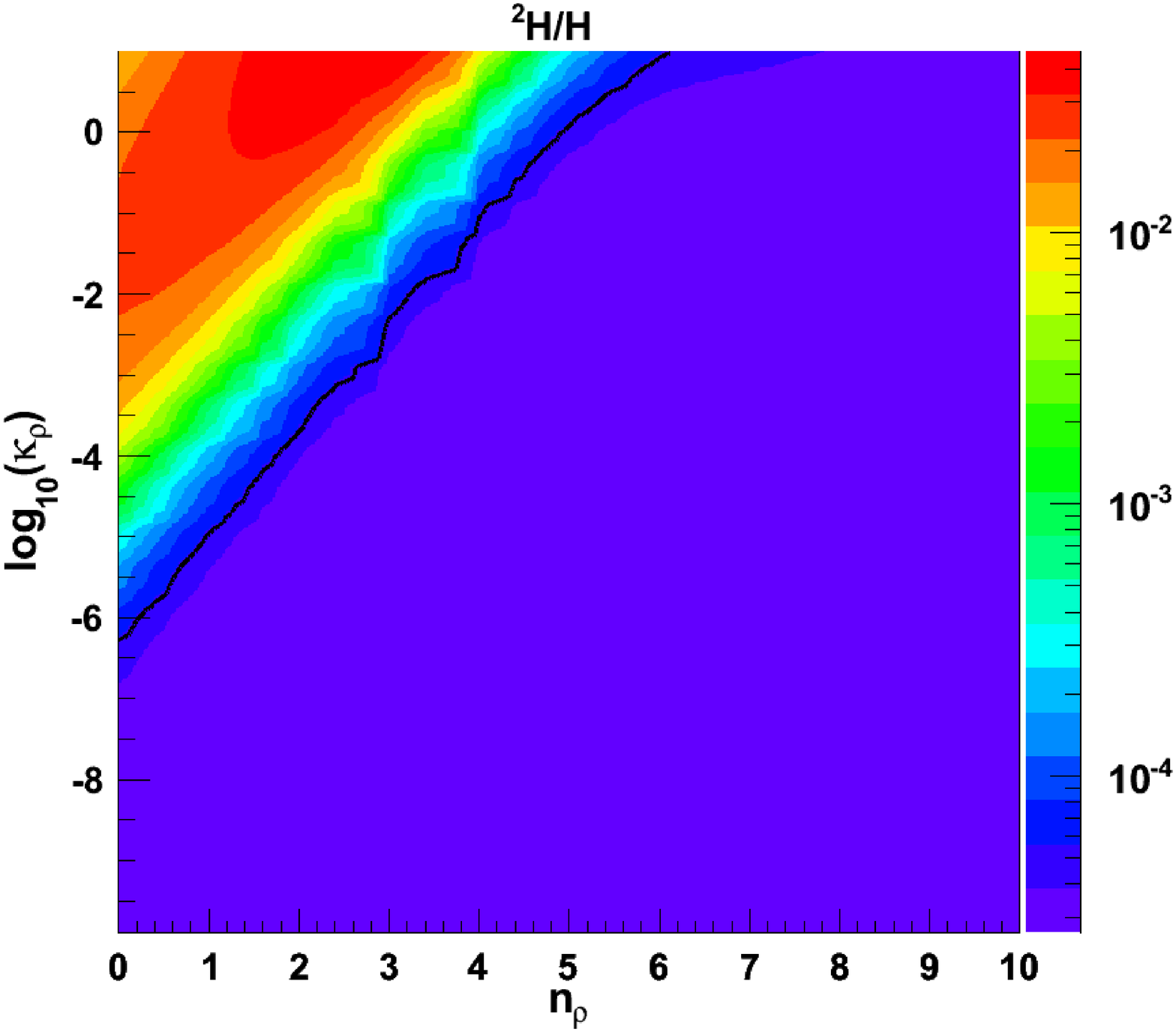}
\caption{Constraints from $Y_p$ (left) and $^2H/H$ (right) on the dark energy parameters $(n_\rho,\kappa_\rho)$. The parameter regions excluded by BBN are located between the black lines for $Y_p$, and in the upper left corner for $^2H/H$. The colors correspond to different values of $Y_p$ and $^2H/H$.\label{BBNenergy}}
\end{figure}%
\begin{figure}[t!]
\includegraphics[width=6.05cm]{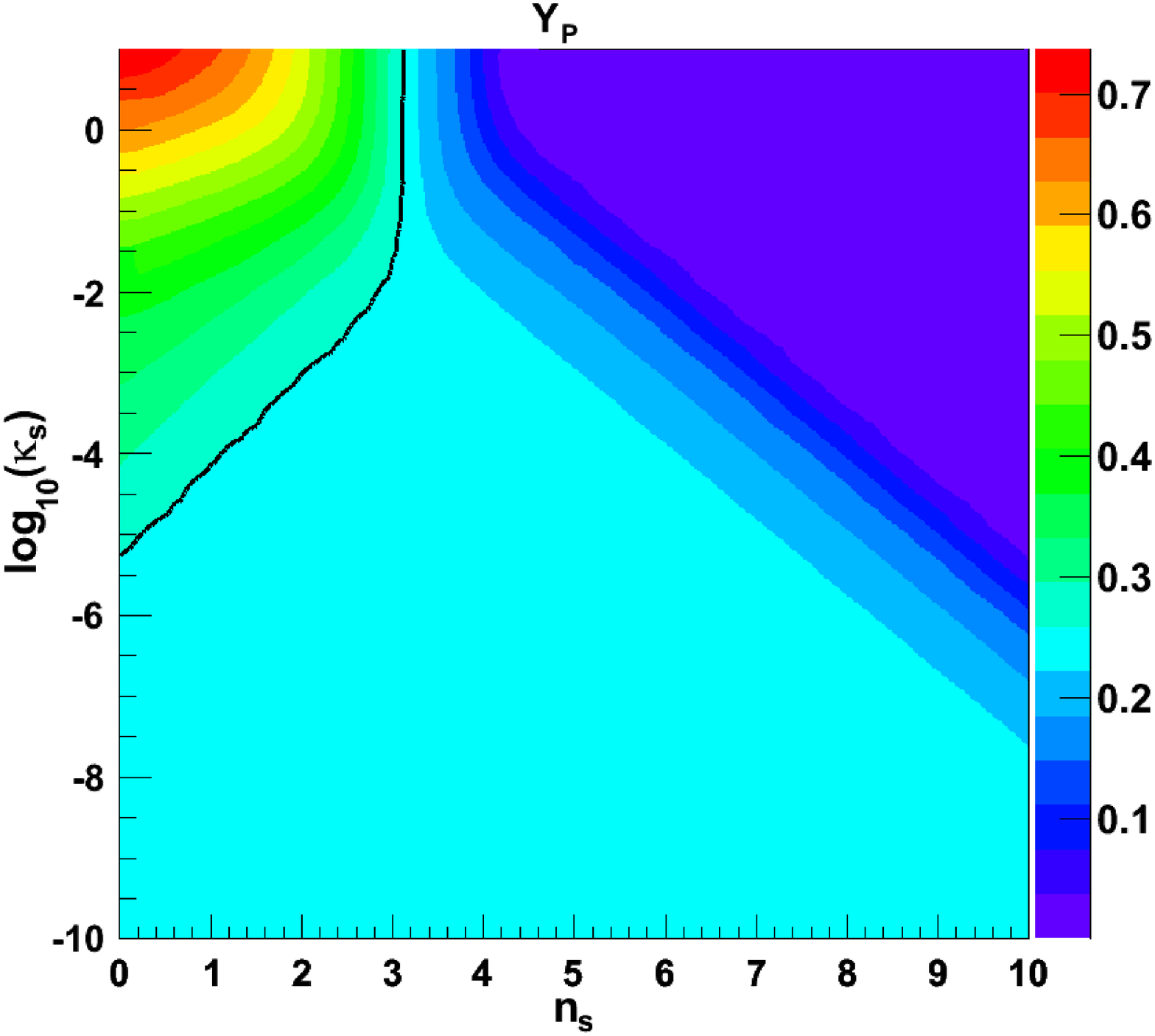}\includegraphics[width=6.05cm]{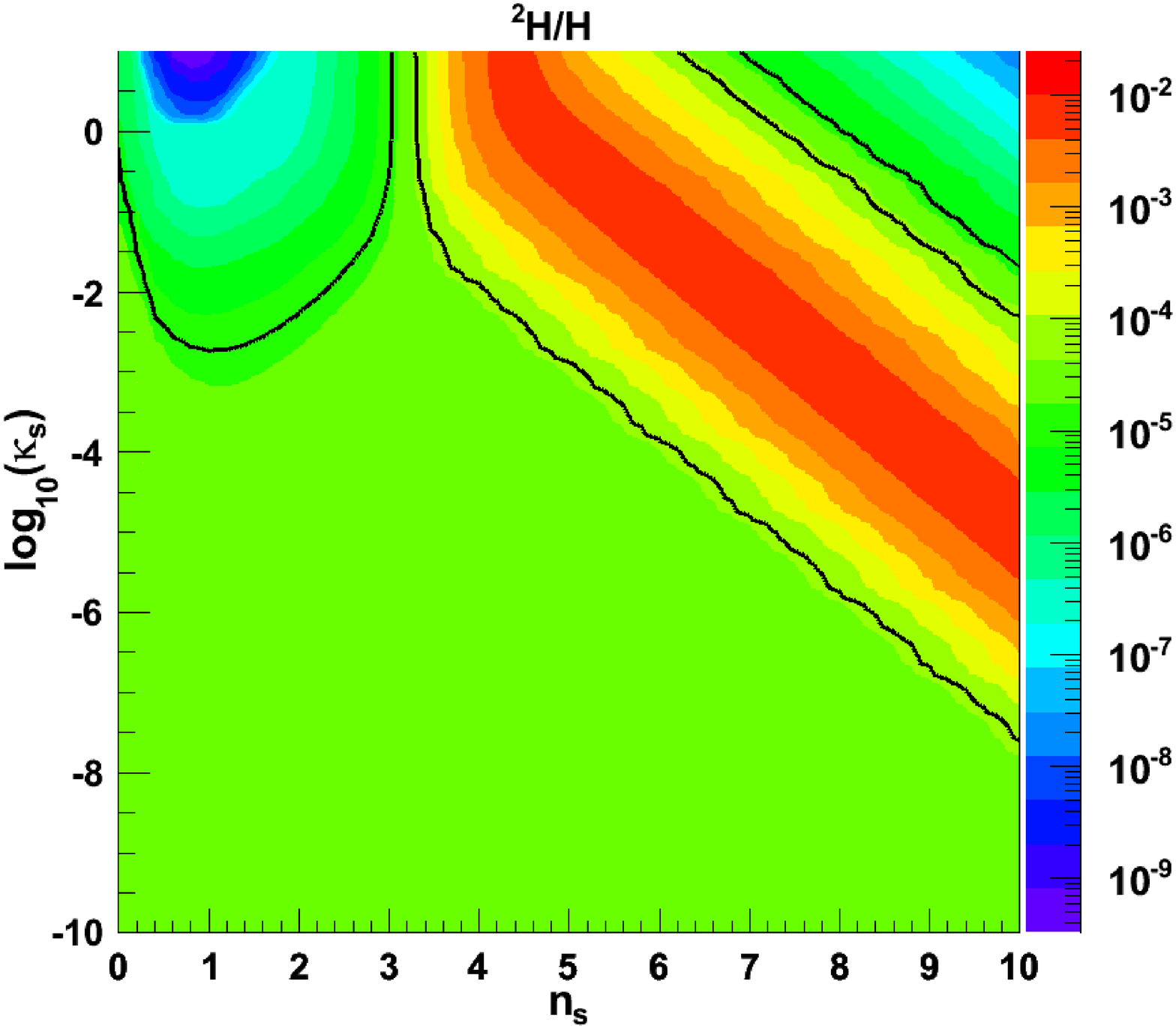}
\caption{Constraints from $Y_p$ (left) and $^2H/H$ (right) on the dark entropy parameters $(n_s,\kappa_s)$. The parameter regions excluded by BBN are located in the upper left corner for $Y_p$, and in both upper corners and in between the right black lines for $^2H/H$. The colors correspond to different values of $Y_p$ and $^2H/H$.\label{BBNentropy}}
\end{figure}%
In Figs.~\ref{BBNenergy} and \ref{BBNentropy} the current limits on the dark energy and entropy properties from the $Y_p$ and $^2H/H$ BBN constraints are presented. The areas on the top of the black lines lead to unfavored element abundances. Comparing Fig. \ref{degeneracy} with Figs. \ref{BBNenergy} and \ref{BBNentropy}, we notice that the left plot ($\kappa_\rho=10^{-2}$, $n_\rho=6$, $\kappa_s=10^{-2}$, $n_s=5$) is disfavored by BBN, while the right one is still compatible. Thus, in such cases, knowing the particle physics properties will enable us to recognize whether the favored cosmological model is the standard one, and by considering also the BBN data, we can already distinguish between different cosmological scenarios.

\section{SuperIso Relic}

The results presented here are obtained using the public code SuperIso Relic v2.7 \cite{Arbey:2009gu}, available on \verb?http://superiso.in2p3.fr/relic? . SuperIso Relic is an extension of the SuperIso program \cite{superiso}, which is devoted to the calculation of flavor physics observables in the Two-Higgs Doublet Model, Minimal Supersymmetric Standard Model and Next-to-Minimal Supersymmetric Standard Model. The main purpose of SuperIso Relic is to compute the relic density in the cosmological standard model as well as in alternative scenarios. All models described here are already implemented, and other models will soon be included. For more information on SuperIso Relic, we refer to its website, or to the manual \cite{Arbey:2009gu}.

\section{Conclusion}
We showed that using the relic density to constrain the supersymmetric parameter space is highly dependent on the cosmological assumptions, and it is then critical to use the relic density constraints for the scientific preparation of the future colliders. We also noticed that inverting the problem, {\it i.e.} using the particle physics results to determine the early Universe properties, can be of great interest as it will give access to an unknown part of the Universe history.


\bibliographystyle{aipproc}   


\begin{thebibliography}{23}
\expandafter\ifx\csname natexlab\endcsname\relax\def\natexlab#1{#1}\fi
\providecommand{\enquote}[1]{``#1''}
\expandafter\ifx\csname url\endcsname\relax
  \def\url#1{\texttt{#1}}\fi
\expandafter\ifx\csname urlprefix\endcsname\relax\def\urlprefix{URL }\fi
\providecommand{\eprint}[2][]{\url{#2}}

\bibitem[Komatsu et~al.(2009)]{Komatsu:2008hk}
E.~Komatsu, et~al., \emph{Astrophys. J. Suppl.} \textbf{180}, 330--376 (2009),
  \eprint{arXiv:0803.0547}.

\bibitem[Burles and Tytler(1998{\natexlab{a}})]{Burles:1997ez}
S.~Burles, and D.~Tytler, \emph{Astrophys. J.} \textbf{499}, 699
  (1998{\natexlab{a}}), \eprint{astro-ph/9712108}.

\bibitem[Burles and Tytler(1998{\natexlab{b}})]{Burles:1997fa}
S.~Burles, and D.~Tytler, \emph{Astrophys. J.} \textbf{507}, 732--744
  (1998{\natexlab{b}}), \eprint{astro-ph/9712109}.

\bibitem[Arbey and Mahmoudi(2008)]{Arbey:2008kv}
A.~Arbey, and F.~Mahmoudi, \emph{Phys. Lett.} \textbf{B669}, 46--51 (2008),
  \eprint{arXiv:0803.0741}.

\bibitem[Ellis et~al.(1997)]{Ellis:1997wva}
J.~R. Ellis, T.~Falk, K.~A. Olive, and M.~Schmitt, \emph{Phys. Lett.}
  \textbf{B413}, 355--364 (1997), \eprint{hep-ph/9705444}.

\bibitem[Gondolo and Gelmini(1991)]{Gondolo:1990dk}
P.~Gondolo, and G.~Gelmini, \emph{Nucl. Phys.} \textbf{B360}, 145--179 (1991).

\bibitem[Edsjo and Gondolo(1997)]{Edsjo:1997bg}
J.~Edsjo, and P.~Gondolo, \emph{Phys. Rev.} \textbf{D56}, 1879--1894 (1997),
  \eprint{hep-ph/9704361}.

\bibitem[Belanger et~al.(2007)]{Belanger:2006is}
G.~Belanger, F.~Boudjema, A.~Pukhov, and A.~Semenov, \emph{Comput. Phys.
  Commun.} \textbf{176}, 367--382 (2007), \eprint{hep-ph/0607059}.

\bibitem[Gondolo et~al.(2004)]{Gondolo:2004sc}
P.~Gondolo, et~al., \emph{JCAP} \textbf{0407}, 008 (2004),
  \eprint{astro-ph/0406204}.

\bibitem[Arbey and Mahmoudi(2009{\natexlab{a}})]{Arbey:2009gu}
A.~Arbey, and F.~Mahmoudi  (2009{\natexlab{a}}), \eprint{arXiv:0906.0369}.

\bibitem[Kamionkowski and Turner(1990)]{Kamionkowski:1990ni}
M.~Kamionkowski, and M.~S. Turner, \emph{Phys. Rev.} \textbf{D42}, 3310--3320
  (1990).

\bibitem[Salati(2003)]{Salati:2002md}
P.~Salati, \emph{Phys. Lett.} \textbf{B571}, 121--131 (2003),
  \eprint{astro-ph/0207396}.

\bibitem[Chung et~al.(2007)]{Chung:2007cn}
D.~J.~H. Chung, L.~L. Everett, K.~Kong, and K.~T. Matchev, \emph{JHEP}
  \textbf{10}, 016 (2007), \eprint{arXiv:0706.2375}.

\bibitem[Moroi and Randall(2000)]{Moroi:1999zb}
T.~Moroi, and L.~Randall, \emph{Nucl. Phys.} \textbf{B570}, 455--472 (2000),
  \eprint{hep-ph/9906527}.

\bibitem[Giudice et~al.(2001)]{Giudice:2000ex}
G.~F. Giudice, E.~W. Kolb, and A.~Riotto, \emph{Phys. Rev.} \textbf{D64},
  023508 (2001), \eprint{hep-ph/0005123}.

\bibitem[Gelmini and Gondolo(2006)]{Gelmini:2006pw}
G.~B. Gelmini, and P.~Gondolo, \emph{Phys. Rev.} \textbf{D74}, 023510 (2006),
  \eprint{hep-ph/0602230}.

\bibitem[Arbey and Mahmoudi(2009{\natexlab{b}})]{Arbey:2009gt}
A.~Arbey, and F.~Mahmoudi  (2009{\natexlab{b}}), \eprint{arXiv:0906.0368}.

\bibitem[Battaglia et~al.(2004)]{Battaglia:2003ab}
M.~Battaglia, et~al., \emph{Eur. Phys. J.} \textbf{C33}, 273--296 (2004),
  \eprint{hep-ph/0306219}.

\bibitem[Ellis et~al.(2007)]{Ellis:2007ka}
J.~Ellis, T.~Hahn, S.~Heinemeyer, K.~A. Olive, and G.~Weiglein, \emph{JHEP}
  \textbf{10}, 092 (2007), \eprint{arXiv:0709.0098}.

\bibitem{softsusy} B.~C. Allanach, \emph{Comput. Phys. Commun.} \textbf{143}, 305 (2002), \eprint{hep-ph/0104145}.

\bibitem[Baltz et~al.(2006)]{Baltz:2006fm}
E.~A. Baltz, M.~Battaglia, M.~E. Peskin, and T.~Wizansky, \emph{Phys. Rev.}
  \textbf{D74}, 103521 (2006), \eprint{hep-ph/0602187}.

\bibitem{superiso}
F.~Mahmoudi, \emph{Comput. Phys. Commun.} \textbf{178}, 745--754 (2008),
  \eprint{arXiv:0710.2067}; F.~Mahmoudi, \emph{Comput. Phys. Commun.} \textbf{180}, 1579--1613
  (2009{\natexlab{a}}), \eprint{arXiv:0808.3144}; F.~Mahmoudi, \emph{Comput. Phys. Commun.} \textbf{180}, 1718--1719
  (2009{\natexlab{b}}).

\end{thebibliography}

\hyphenation{Post-Script Sprin-ger}

\end{document}